\documentclass[11pt,reqno,oneside]{amsart}
\usepackage{amssymb}
\usepackage[reqno]{amsmath}

\usepackage{graphicx}
\usepackage{subfigure}

\numberwithin{equation}{section}

\begin{document}

\input epsf     

\begin{flushright}
\baselineskip=12pt
CERN--TH/98--35,\quad UTPT--98--01\\
\end{flushright}

\title[Dynamical Instability of the Wyman Solution]{The Dynamical Instability of Static, Spherically Symmetric Solutions in Nonsymmetric Gravitational Theories}
\author[1]{M.~A. Clayton}
\author[2]{L. Demopoulos}
\author[3]{J. L\'{e}gar\'{e}}

\address[1]{CERN--Theory Division, CH--1211 Geneva 23, Switzerland}
\email[1]{Michael.A.Clayton@cern.ch}

\address[2,3]{Department of Physics, University of Toronto, Toronto, \textsc{on}, Canada M5S
1A7}
\email[2]{terry@medb.physics.utoronto.ca}
\email[3]{jacques@medb.physics.utoronto.ca}

\date{\today}

\begin{abstract}
We consider the dynamical stability of a class of static, spherically-symmetric solutions of the nonsymmetric gravitational theory.
We numerically reproduce the Wyman solution and generate new solutions for the case where the theory has a nontrivial fundamental length scale $\mu^{-1}$.
By considering spherically symmetric perturbations of these solutions we show that the Wyman solutions are generically unstable.
\end{abstract}

\maketitle

\section{Introduction}
\label{sect:Intro}

The goal of this work is to understand the role of a class of static, spherically-symmetric solutions of the nonsymmetric gravitational theory~\cite{Moffat:1979,Moffat:1990} (NGT), which are also solutions to Einstein's unified field theory~\cite{Einstein+Straus:1946} (UFT).
In Section~\ref{sect:general}, we show that the time-dependent field equations for this sector are equivalent to an E-KG system with metric $\hat{\mathrm{g}}$ and a scalar field $\psi$ with a positive semi-definite (coordinate-dependent) self-interaction potential.
This result allows us to use a scaling argument to show in Section~\ref{sect:background} that there are neither solutions with nontrivial $\psi$ and globally regular $\hat{\mathrm{g}}$, nor solutions with an event horizon and nonzero $\psi$ in the exterior (no hair).
The remaining class of static solutions (regular and no event horizons outside of the origin) consist of the Wyman solution~\cite{Wyman:1950,Bonnor:1951} and generalizations to solutions of the theory that possesses a non-zero inverse length scale $\mu$, both of which are numerically generated in Section~\ref{sect:numerical background}.

The issue at hand is whether these solutions may be considered as a possible endpoint of gravitational collapse in NGT, a matter which has been discussed in the literature~\cite{Burko+Ori:1995,Dobrowolski+Koc:1996,Moffat:1995,Moffat+Sokolov:1995b}.
To answer this question we investigate the dynamical stability of these solutions under spherically symmetric perturbations.
Employing both a variational argument in Section~\ref{sect:variation} as well as determining the ground state and its eigenvalue directly, in Section~\ref{sect:Dynamical Inst} we find that these solutions are dynamically unstable.
The instability of the $\mu=0$ case is analogous to that of an Einstein-Klein Gordon system (E-KG) where it was found that the static solutions with nontrivial scalar field (also attributed to Wyman~\cite{Wyman:1981}) are unstable~\cite{Jetzer+Scialom:1992}.

\section{The Wyman Sector Field Equations}
\label{sect:general}

Here we will give the (coordinate frame) action for NGT as presented in~\cite{Clayton:1995}
\begin{equation}\label{eq:NGTAction}
S_{\textsc{ngt}}=\int d^4\!x\,\sqrt{-\mathrm{g}}\Bigl(
-\mathrm{g}^{\mu\nu}R_{\mu\nu}^{\textsc{ns}}
-\mathrm{g}^{\mu\nu}\nabla_{[\mu}[W]_{\nu]}
+\tfrac{1}{2}\alpha\mathrm{g}^{\mu\nu}W_\mu W_\nu
+l^\mu\Lambda_\mu
+\tfrac{1}{4}\mu^2\mathrm{g}^{[\mu\nu]}\mathrm{g}_{[\mu\nu]}\Bigr).
\end{equation}
The inverse of the fundamental tensor is defined by 
$\mathrm{g}^{\mu\gamma}\mathrm{g}_{\gamma\nu}=\mathrm{g}_{\nu\gamma}\mathrm{g}
^{\gamma\mu}=\delta^\mu_\nu$, and $\mathrm{g}:=\det[\mathrm{g}_{\mu\nu}]$.
The covariant derivative is characterized by the torsion-free ($\Gamma^\alpha_{[\mu\nu]}=0$) connection coefficients 
$\nabla_\mu[\partial_\nu]=\Gamma^\gamma_{\mu\nu}\partial_\gamma$, not required to be compatible with any tensor.
What are commonly treated as the antisymmetric components of the connection coefficients are considered as an additional tensor $\Lambda^\gamma_{\mu\nu}$.

The Ricci-like tensor that appears in the action is split into two contributions $R_{\mu\nu}^{\textsc{ns}}=R_{\mu\nu}+R^\Lambda_{\mu\nu}$; the first is identified as the Ricci tensor defined from $\Gamma^\gamma_{\mu\nu}$ (and reduces to the GR Ricci tensor in the limit of vanishing antisymmetric sector), and the second contains contributions from the antisymmetric tensor field $\Lambda^\gamma_{\mu\nu}$ ($\Lambda_\mu:=\Lambda^\gamma_{\mu\gamma}$).
These are:
\begin{subequations}
\begin{align}
\label{eq:R}
R_{\mu\nu}=&\partial_\gamma[\Gamma^\gamma_{\nu\mu}]
-\tfrac{1}{2}\partial_\nu[\Gamma^\gamma_{\gamma\mu}]
-\tfrac{1}{2}\partial_\mu[\Gamma^\gamma_{\nu\gamma}]
+\Gamma^\delta_{\nu\mu}\Gamma^\gamma_{\gamma\delta}
-\Gamma^\delta_{\gamma\nu}\Gamma^\gamma_{\delta\mu},\\
\label{eq:R Lam}
R^\Lambda_{\mu\nu}=&
\nabla_\gamma[\Lambda]^\gamma_{\mu\nu}
+\nabla_{[\mu}[\Lambda]_{\nu]}
+\Lambda^\gamma_{\mu\delta}\Lambda^\delta_{\nu \gamma}.
\end{align}
\end{subequations}
In addition to the vanishing of the torsion tensor, the compatibility 
conditions of NGT are~\cite{Clayton:1995}
\begin{subequations}\label{eq:comp eqns}
\begin{gather}
\Lambda_\mu=0, \quad
\nabla_\nu[\sqrt{-\mathrm{g}}\mathrm{g}]^{[\mu\nu]}
=\alpha\sqrt{-\mathrm{g}}\mathrm{g}^{(\mu\nu)}W_\nu,\quad
l^\mu=\tfrac{1}{3}\alpha\mathrm{g}^{(\mu\nu)}W_\nu,\\
\label{eq:compat}
\nabla_{\alpha}[\mathrm{g}]_{\mu\nu}
=\mathrm{g}_{\mu\gamma}\Lambda^\gamma_{\nu\alpha}
+\mathrm{g}_{\gamma\nu}\Lambda^\gamma_{\alpha\mu}
+\tfrac{2}{3}\alpha\bigl(
\mathrm{g}_{\mu[\alpha}\mathrm{g}_{\gamma]\nu}
+\tfrac{1}{2}\mathrm{g}_{\mu\nu}\mathrm{g}_{[\alpha\gamma]}
\bigr)\mathrm{g}^{(\gamma\delta)}W_\delta,
\end{gather}
\end{subequations}
and the field equations are given by
\begin{equation}
\label{eq:mNGTF}
\mathcal{R}_{\mu\nu}=R^{\textsc{ns}}_{\mu\nu}
+\partial_{[\mu}[W]_{\nu]}
-\tfrac{1}{2}\alpha W_\mu W_\nu
-\tfrac{1}{4}\mu^2M_{\mu\nu}=0, 
\end{equation}
where
\begin{equation}\label{eq:mass tensor}
M_{\mu\nu}=\mathrm{g}_{[\mu\nu]}
-\mathrm{g}_{\gamma\mu}\mathrm{g}_{\nu\delta}\mathrm{g}^{[\gamma\delta]}
+\tfrac{1}{2}\mathrm{g}_{\nu\mu}
\mathrm{g}^{[\gamma\delta]}\mathrm{g}_{[\gamma\delta]}.
\end{equation}
The action (\ref{eq:NGTAction}) and field equations~\eqref{eq:comp eqns} and~\eqref{eq:mNGTF} encompass those of the `massive' theory~\cite{Legare+Moffat:1995,Clayton:1995,Moffat:1994,Moffat:1995b} when $\alpha=3/4$, `old' NGT~\cite{Moffat:1990} with vanishing source current for $\alpha=0$ and $\mu=0$ (which is equivalent to UFT~\cite{Einstein:1945,Einstein+Straus:1946}), and recovers GR in the limit that all antisymmetric components of the fundamental tensor are set to zero~\cite{Clayton:1995,Clayton:Phd:1996}.

The general form of the spherically-symmetric fundamental tensor~\cite{Papapetrou:1948} consists of the general form of a spherically symmetric metric with the additional antisymmetric components $\mathrm{g}_{[01]}=\omega(t,r)$ and $\mathrm{g}_{[23]}=f(t,r)\sin(\theta)$; in this work we consider the Wyman sector~\cite{Wyman:1950,Bonnor:1951} defined by choosing $\mathrm{g}_{[01]}=0$.
The remaining components will be parameterised by the symmetric sector functions $\nu$ and $\lambda$, and a dimensionless field $\psi$ parameterising the mixing of the angular components
\begin{equation}\label{eq:ft param}
\lvert\mathrm{g}_{(\mu\nu)}\rvert
=\mathrm{diag}\bigl(\mathrm{e}^\nu,-\mathrm{e}^\lambda,
-r^2\cos(\psi),-r^2\cos(\psi)\sin^2(\theta)\bigr),\quad
\mathrm{g}_{[23]}=r^2\sin(\psi)\sin(\theta).
\end{equation}
From this form of the fundamental tensor one finds the following combinations of the field equations~\eqref{eq:mNGTF} which will prove useful:
\begin{subequations}\label{eq:feq used}
\begin{align}
\mathcal{R}_{01}=&\tfrac{1}{r}\bigl\{\partial_t[\lambda]
-\tfrac{r}{2}\partial_t[\psi]\partial_r[\psi]\bigr\},\\
\mathrm{e}^{-\nu}\mathcal{R}_{00}+\mathrm{e}^{-\lambda}\mathcal{R}_{11}=&
\tfrac{1}{r}\mathrm{e}^{-\lambda}\bigl\{
\partial_r[\nu+\lambda]
-\tfrac{r}{2}\mathrm{e}^{\lambda-\nu}(\partial_t[\psi])^2
-\tfrac{r}{2}(\partial_r[\psi])^2\bigr\},\\
-\tfrac{1}{r^2}\bigl(\cos(\psi)\mathcal{R}_{22}
+\sin(\psi)\mathcal{R}_{23}/\sin(\theta)\bigr)=&\nonumber \\
\tfrac{1}{2r}\mathrm{e}^{-\lambda}\bigl\{
\partial_r[\nu-\lambda]
+{}&\tfrac{2}{r}\bigl(1-\mathrm{e}^\lambda\cos(\psi)\bigr)
+\tfrac{1}{2}\mu^2r\mathrm{e}^\lambda\sin^2(\psi)\bigr\},\\
-\tfrac{1}{r^2}\bigl(\sin(\psi)\mathcal{R}_{22}
-\cos(\psi)\mathcal{R}_{23}/\sin(\theta)\bigr)=&
\tfrac{1}{2}\mathrm{e}^{-\lambda}\bigl\{
-\mathrm{e}^{\lambda-\nu}\partial_t^2[\psi]
+\tfrac{1}{2}\mathrm{e}^{\lambda-\nu}\partial_t[\nu-\lambda]\partial_t[\psi]
\nonumber \\
\label{eq:wave psi}
{}+\partial^2_r[\psi]
+\tfrac{1}{2}\partial_r[\nu-\lambda]\partial_r[\psi]
+{}&\tfrac{2}{r}\partial_r[\psi]
-\tfrac{2}{r^2}\mathrm{e}^\lambda\sin(\psi)
-\tfrac{1}{2}\mathrm{e}^\lambda\mu^2\sin(2\psi)\bigr\}.
\end{align}
\end{subequations}

Examination of the field equations~\eqref{eq:feq used} reveals a strong similarity to an E-KG system.
Pursuing this analogy we introduce the Riemannian metric
\begin{equation}\label{eq:ghat}
\hat{\mathrm{g}}=\mathrm{diag}\bigl(\mathrm{e}^\nu,-\mathrm{e}^\lambda,
-r^2,-r^2\sin^2(\theta)\bigr).
\end{equation}
Denoting the Ricci tensor derived from $\hat{\mathrm{g}}$ as $\hat{R}_{\mu\nu}$, the field equations may be written in the E-KG form:
\begin{equation}\label{eq:hat equations}
\begin{split}
\hat{R}_{00}=&\tfrac{1}{2}(\partial_t[\psi])^2
-\tfrac{1}{4}\mu^2\mathrm{e}^{\nu}\sin^2(\psi),\quad
\hat{R}_{01}=\tfrac{1}{2}\partial_t[\psi]\partial_r[\psi],\\
\hat{R}_{11}=&\tfrac{1}{2}(\partial_r[\psi])^2
+\tfrac{1}{4}\mu^2\mathrm{e}^{\lambda}\sin^2(\psi),\quad
\hat{R}_{22}=1-\cos(\psi)+\tfrac{1}{4}\mu^2r^2\sin^2(\psi).
\end{split}
\end{equation}
Additionally we have the wave equation~\eqref{eq:wave psi} for $\psi$, which may be written
as
\begin{equation}\label{eq:waves}
\hat{\mathrm{g}}^{\mu\nu}\hat{\nabla}_\mu\hat{\nabla}_\nu[\psi]
+\delta_\psi[\hat{V}]=0,
\end{equation}
where $\delta_\psi[\hat{V}]$ is the functional derivative with respect to $\psi$ of the potential defined by
\begin{equation}\label{eq:Wyman potential}
\hat{V}:=\tfrac{2}{r^2}\bigl(1-\cos(\psi)\bigr)
+\tfrac{1}{2}\mu^2\sin^2(\psi).
\end{equation}
Although the field equations for this sector have appeared in various places~\cite{Tonnelat:1982,Pant:1975,Moffat:1995}, to date the analogy with an Einstein-Klein Gordon system has not been realised.
This identification is useful since the E-KG system is known to possess a unique light cone determined by the metric $\hat{\mathrm{g}}$, and allows us to identify the metric $\hat{\mathrm{g}}$ (up to a confirmal factor) as a physical measure of spacetime.
This is a nontrivial statement in the UFT in general given the existence of multiple light cones~\cite{Maurer-Tison:1959}. 
It is also straightforward to show that if~\eqref{eq:hat equations} are written as $\hat{G}_{\mu\nu}=\tfrac{1}{2}\hat{T}_{\mu\nu}$, then the Bianchi identity $\hat{\nabla}_\nu[\hat{G}]^\nu_{\;\mu}=0$ is satisfied by $\hat{T}_{\mu\nu}$ by virtue of the wave equation~\eqref{eq:waves}.
Thus although the Bianchi identities for the nonsymmetric theory~\cite{Legare+Moffat:1995} do not in general imply separate conservation laws for a general relativistic sector and an antisymmetric sector written as a simple matter field, for the Wyman sector this is realised.

This provides some motivation for studying this sector; considered on its own it suffers from neither the bad asymptotic behaviour noted in~\cite{Damour+Deser+McCarthy:1992,Damour+Deser+McCarthy:1993} nor the linearization instability discussed in~\cite{Clayton:1996}, and is reproduced by many of the nonsymmetric field actions.
Thus one conjectures that when an acceptable nonsymmetric action is found, it will also reproduce this sector.
We show that if we identify $\hat{\mathrm{g}}$ with the metric of an E-KG system, the static, spherically-symmetric Wyman spacetimes with nonvanishing $\psi$ and the E-KG spacetimes with nonvanishing scalar field (also attributed to Wyman~\cite{Wyman:1981,Schmoltzi+Schucker:1991}) have similar properties.
Thus we expect that they have similar dynamical stability properties~\cite{Jetzer+Scialom:1992,Choptuik+Hirschmann+Liebling:1997}, a result that we will establish below.

\section{The Static Background Spacetimes}
\label{sect:background}

We now turn to the task of generating the static background fields $\nu_0$, $\lambda_0$, and $\psi_0$.
Since we are going to show that these solutions are dynamically unstable, it is worthwhile to include a proof that the solutions for which $\hat{\mathrm{g}}$ is globally regular or has an event horizon with regular exterior consist precisely of the Minkowski spacetime and the Schwarzschild solution.
The argument is an extension of that given by Heusler~\cite{Heusler:1996}.

Writing $\mathrm{e}^\nu=n^2(r)\bigl(1-2m(r)/r\bigr)$ and $\mathrm{e}^\lambda=1/\bigl(1-2m(r)/r\bigr)$ and inserting the fundamental tensor~\eqref{eq:ft param} into the Lagrangian density~\eqref{eq:NGTAction} we find the effective action for static configurations (using $\int d\Omega=4\pi$)
\begin{equation}\label{eq:eff action}
S^{\text{eff}}=4\pi \int_{b}^\infty dr\,
n\Bigl[4\partial_r[m]
-\frac{1}{2}r^2\Bigl(1-\frac{2m}{r}\Bigr)\bigl(\partial_r[\psi]\bigr)^2
-2\bigl(1-\cos(\psi)\bigr)
-\frac{1}{2}r^2\mu^2\sin^2(\psi)\Bigr].
\end{equation}
It is straightforward to show that variations of this action with respect to $n(r)$, $m(r)$ and $\psi(r)$ are equivalent to the static limit of~\eqref{eq:feq used}.

Assuming that we have an asymptotically flat solution to the static field equations
\begin{equation}
m\xrightarrow{r\rightarrow\infty}M_{\textsc{s}}+o(1/r),\quad
n\xrightarrow{r\rightarrow\infty}1+o(1/r),\quad
\psi\xrightarrow{r\rightarrow\infty}o(1/r),
\end{equation}
with either a regular centre at $r=b=0$ so that
\begin{subequations}\label{eq:bcs}
\begin{equation}
\lim_{r\rightarrow 0}\psi \text{ is finite},\quad
\lim_{r\rightarrow 0}n \text{ is finite},\quad
m\sim m_0 r +\mathcal{O}(r^2)\text{ with } m_0< 1/2, 
\end{equation}
or an event horizon at $r=b$ where
\begin{equation}
\lim_{r\rightarrow b}\psi \text{ is finite},\quad
\lim_{r\rightarrow b}n \text{ is finite},\quad
\lim_{r\rightarrow b}m=b/2,
\end{equation}
\end{subequations}
then the one-parameter family of fields $n_\lambda$, $m_\lambda$, and $\psi_\lambda$, where for example $m_\lambda(r):=m\bigl((1-\lambda)b+\lambda r\bigr)$, have the same boundary values.
The action $S^{\text{eff}}_\lambda:=S^{\text{eff}}[n_\lambda, m_\lambda, \psi_\lambda]$ must therefore have a critical point at $\lambda=1$, and so $\partial_\lambda\bigl[S^{\text{eff}}_\lambda\bigr]\rvert_{\lambda=1}=0$.
Performing this variation on~\eqref{eq:eff action} results in
\begin{multline}\label{eq:eff var}
\partial_\lambda\bigl[S^{\text{eff}}_\lambda\bigr]\big{\rvert}_{\lambda=1}=
4\pi \int_b^\infty dr\;n\Bigl[
\tfrac{1}{2}\bigl(r(r-b)-b(r-2m)\bigr)\bigl(\partial_r[\psi]\bigr)^2\\
+2\bigl(1-\cos(\psi)\bigr)
+\tfrac{1}{2}\bigl(r^2+2r(r-b)\bigr)\mu^2\sin^2(\psi)\Bigr]=0.
\end{multline}
The final two terms are manifestly positive semi-definite, and the first may be seen to be as well by the following argument: the coefficient of $(\partial_r[\psi])^2$ vanishes at $r=b$ for either a regular centre or a horizon, and its derivative $2(r-b)+2b\partial_r[m]$ is non-negative for $r>b$ (this follows from the variation of~\eqref{eq:eff action} with respect to $n(r)$, which shows that $\partial_r[m]>0$).
Therefore all terms in~\eqref{eq:eff var} must vanish separately and we have shown that no solution exists other than the trivial one $\psi=0$.
That these results are a straightforward extension of those for an E-KG system as derived by Heusler is due to the positive semi-definiteness of the potential~\eqref{eq:Wyman potential}.

We therefore know that static solutions to the field equations that have nontrivial antisymmetric components of the fundamental tensor must have neither event horizon nor be globally regular.
It is to the numerical determination of such solutions that we turn next.

\subsection{Numerical Results}
\label{sect:numerical background}

Throughout the remainder of this work we will make use of the dimensionless radial coordinate
scaled by the asymptotic Schwarzschild mass parameter $M_{\textsc{s}}$ of the system:
$x:=r/(2M_{\textsc{s}})$.
In the case of a non-zero length scale $\mu$ we will also make use of this coordinate, additionally defining $\tilde{\mu}:=2M_{\textsc{s}}\mu$.
We then re-write the static limit of~\eqref{eq:feq used} in a form appropriate for the numerical integration implemented below:
\begin{subequations}\label{eq:num eqns}
\begin{align}
\partial_x[\mathrm{e}^{\nu_0}]=&
\mathrm{e}^{\nu_0}\bigl\{
\tfrac{x}{4}(\partial_x[\psi_0])^2
-\tfrac{1}{x}\bigl(1-\mathrm{e}^{\lambda_0}\cos(\psi_0)\bigr)
-\tfrac{1}{4}\tilde{\mu}^2x\mathrm{e}^{\lambda_0}\sin^2(\psi_0)\bigr\},\\
\partial_x[\mathrm{e}^{\lambda_0}]=&
\mathrm{e}^{\lambda_0}\bigl\{
\tfrac{x}{4}(\partial_x[\psi_0])^2
+\tfrac{1}{x}\bigl(1-\mathrm{e}^{\lambda_0}\cos(\psi_0)\bigr)
+\tfrac{1}{4}\tilde{\mu}^2x\mathrm{e}^{\lambda_0}\sin^2(\psi_0)\bigr\},
\end{align}
\begin{multline}
\label{eq:psi0 equation}
\partial^2_x[\psi_0]
+\tfrac{1}{x}\bigl(1+\mathrm{e}^{\lambda_0}\cos(\psi_0)
-\tfrac{1}{4}\tilde{\mu}^2x^2\mathrm{e}^{\lambda_0}\sin^2(\psi_0)\bigr)
\partial_x[\psi_0] \\
-\tfrac{2}{x^2}\mathrm{e}^{\lambda_0}\sin(\psi_0)
-\tfrac{1}{2}\tilde{\mu}^2\mathrm{e}^{\lambda_0}\sin(2\psi_0)=0.
\end{multline}
\end{subequations}

\begin{figure}[h]
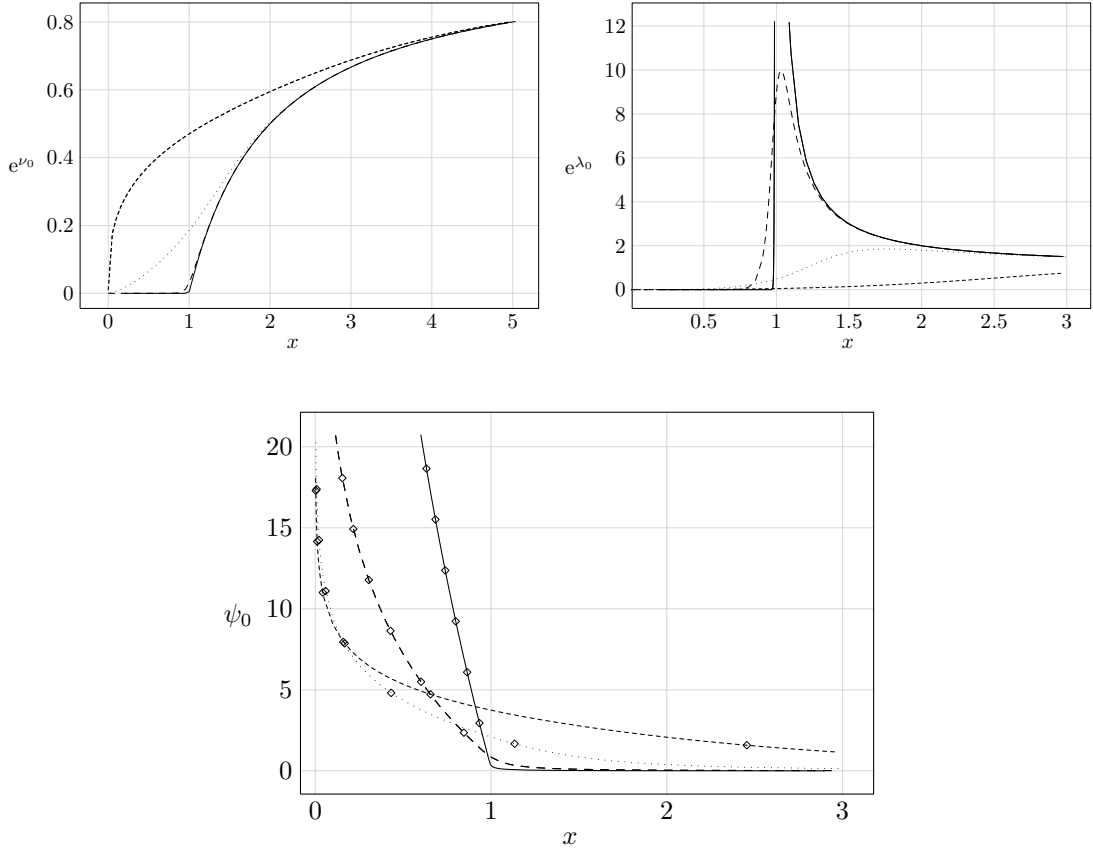

\begin{center}
\begin{subfigure}
{\includegraphics[scale=0.8]{f.ps}}
\end{subfigure}
\begin{subfigure}
{\includegraphics[scale=0.8]{g.ps}}
\end{subfigure}
\begin{subfigure}
{\includegraphics[scale=1.0]{psi.ps}}
\end{subfigure}
\end{center}
\caption{The Wyman solution ($\tilde{\mu}=0$): the solid line corresponds to $s=0.2$, the long-dashed line to $s=1$, the dotted line to $s=10$, and the short-dashed line to  $s=100$.
The first six analytic points from~\eqref{eq:points} are indicated by diamonds on the plot of $\psi_0$.
}
\label{fig:psi}
\end{figure}

We will employ the simple shooting method~\cite[Section~7.3.1]{Stoer+Bulirsch:1993} on $\psi_0$ to numerically generate a solution to~\eqref{eq:num eqns}.
Initial values for $\mathrm{e}^{\nu_0}$, $\mathrm{e}^{\lambda_0}$, $\psi_0$ and $\partial_x[\psi_0]$ are given at a small $x$ and~\eqref{eq:num eqns} are integrated outward using a fourth-order Runge-Kutta~\cite[Section~7.2.1]{Stoer+Bulirsch:1993} routine developed by the authors.
It is therefore necessary to have the small-$x$ behaviour of these solutions, which may be determined from~\eqref{eq:num eqns}.
Assuming that both of $\mathrm{e}^{\nu_0}$ and $\mathrm{e}^{\lambda_0}$ vanish as some positive power of $x$ as $x\rightarrow 0$, we find

\begin{equation}\label{eq:small}
\mathrm{e}^{\nu_0}\sim 
F x_s^{2/(a_p-1)},\quad
\mathrm{e}^{\lambda_0}\sim
G \frac{4\sqrt{1+s^2}}{(a_p-1)^2}
x_s^{2a_p/(a_p-1)},\quad
\psi_0\sim 
-\frac{2a_m}{a_p-1}\ln(x_s)+B,
\end{equation}
independent of the value of $\tilde{\mu}$.
In the $\tilde{\mu}=0$ case we know from~\eqref{eq:small x} that $F=G=1$ and from~\eqref{eq:psi small x} that $B=\tan^{-1}(-s)$.
As we show in Figure~\ref{fig:psi}, the numerical solutions reproduce the analytic results quite well.
By shooting on $B$ we have checked that the shooting algorithm reproduces $B=\tan^{-1}(-s)$ to machine accuracy.

The large-$x$ behaviour of the solutions is determined by requiring that the metric components behave asymptotically like the Schwarzschild solution:
\begin{equation}\label{eq:asmpt Sch}
\mathrm{e}^{\nu_0}\sim 1-\tfrac{1}{x},\quad
\mathrm{e}^{\lambda_0}\sim 1+\tfrac{1}{x}.
\end{equation}
Inserting this into~\eqref{eq:psi0 equation} we find (keeping only the asymptotically
dominant terms and noting that we require that $\psi\rightarrow 0$)
\begin{equation}
\partial^2_x[\psi_0]
+\tfrac{2}{x}\partial_x[\psi_0]
-\tfrac{2}{x^2}\psi_0
-\tilde{\mu}^2\psi_0\sim 0.
\end{equation}
From this the asymptotic form of $\psi_0$ is determined:
\begin{equation}
\psi_0\sim\begin{cases}
A/x^2& \text{for }\tilde{\mu}=0\\
A\mathrm{e}^{-\tilde{\mu}x}/x&\text{for }\tilde{\mu}\neq 0
\end{cases},
\end{equation}
and from~\eqref{eq:large x} we know that for $\tilde{\mu}=0$ we must find $A=s/12$, a result that has been verified numerically.

For the $\tilde{\mu}\neq 0$ solutions things become more complicated.
Although the constant $F$ may be freely chosen by rescaling the time variable, we still must determine the two constants $G$ and $B$, which may be shot for by requiring both $\psi_0\rightarrow 0$ and~\eqref{eq:asmpt Sch}.
In order to avoid shooting on two variables we proceed as follows: 
We choose a value of $s$ and $\tilde{\mu}$ and assign $F=G=1$ initially,  then shoot on $\psi_0$ to determine $B$.
This will result in a solution with asymptotic behaviour $\mathrm{e}^{\lambda_0}\sim (1+\alpha/x)$ and $\mathrm{e}^{\nu_0}\sim f_0(1-\alpha/x)$.
Re-scaling $x$ by $x\rightarrow \alpha x$ and renormalizing $\mathrm{e}^{\nu_0}$ results in a solution with the correct asymptotic form for the metric functions, while the $x\sim 0$ forms become 
\begin{equation}
\mathrm{e}^{\nu_0}\sim f_0^{-1} (\alpha x_s)^{2/(a_p-1)}\quad
\mathrm{e}^{\lambda_0}\sim 4\sqrt{1+s^2}(\alpha x_s)^{2a_p/(a_p-1)}/(a_p-1)^2.
\end{equation}
The constant $B$ determined by shooting is shifted by $B\rightarrow B-2a_m\ln(\alpha)/(a_p-1)$ and, most importantly, we have $\tilde{\mu}\rightarrow \alpha\tilde{\mu}$.

\begin{figure}[h]
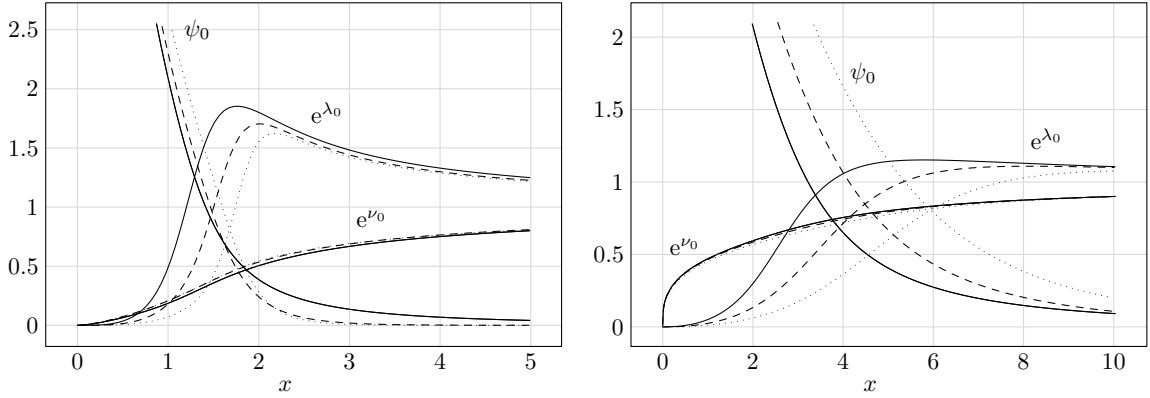

\begin{center}
\begin{subfigure}
[$s=10$: the solid line corresponds to $\tilde{\mu}\approx 0.01001$, 
the dashed line to $\tilde{\mu}\approx 1.551$, 
and the dotted line to the $\tilde{\mu}\approx 2.418$ solution.]
{\includegraphics[scale=0.9]{back-10.0.ps}}
\end{subfigure}
\begin{subfigure}
[$s=100$: the solid line corresponds to $\tilde{\mu}\approx 0.009995$, 
the dashed line to $\tilde{\mu}\approx 0.1471$, 
and the dotted line to the $\tilde{\mu}\approx 0.1625$ solution.]
{\includegraphics[scale=0.9]{back-100.0.ps}}
\end{subfigure}
\end{center}
\caption{A selection of solutions with $\tilde{\mu}\neq 0$.}
\label{fig:mus}
\end{figure}

In Figure~\ref{fig:mus}, we show the numerically generated solutions for various values of $s$ and $\tilde{\mu}$.
We find that initially for increasing values of $\tilde{\mu}$ input into the above rescaling procedure the re-scaled $\tilde{\mu}$ is also increasing, however it then turns over and decreases towards zero.

\section{Dynamical Stability}
\label{sect:Dynamical Inst}

Finally we turn to the investigation of the dynamical stability of these spacetimes, \textit{i.e.}, whether they are unstable against linear perturbations.
Two different methods will be employed: a variational approach which casts the equation for the perturbative modes as the solution of a stationary Schr\"{o}dinger equation and uses a one-parameter family of wave functions to put an upper bound on the lowest eigenvalue, and a shooting method to numerically generate both the lowest energy mode as well as the eigenvalue.

In either case we make use of the single Fourier-mode expansions of the perturbing fields $\nu_1$, $\lambda_1$ and $\psi_1$:
\begin{gather}
\nu(t,x)=\nu_0(x)+\nu_1(x)\cos(\tilde{\omega} t),\quad
\lambda(t,x)=\lambda_0(x)+\lambda_1(x)\cos(\tilde{\omega} t),\\
\psi(t,x)=\psi_0(x)+\tfrac{1}{x}\psi_1(x)\cos(\tilde{\omega} t).\nonumber
\end{gather}
The first-order perturbation equations from~\eqref{eq:feq used} for the symmetric functions
give
\begin{subequations}\label{eq:pert eqns}
\begin{gather}
\lambda_1=\tfrac{1}{2}\partial_x[\psi_0]\psi_1,\quad 
\partial_x[\nu_1]=-\tfrac{1}{2}\partial^2_x[\psi_0]\psi_1
-\tfrac{1}{x}\partial_x[\psi_0]\psi_1
+\tfrac{1}{2}\partial_x[\psi_0]\partial_x[\psi_1],
\end{gather}
and the equation for $\psi_1$
\begin{equation}\label{eq:psi 1}
\partial_x^2[\psi_1]
+\tfrac{1}{2}\partial_x[\nu_0-\lambda_0]\partial_x[\psi_1]
-\mathrm{e}^{\lambda_0-\nu_0}V[x]\psi_1
+\tilde{\omega}^2\mathrm{e}^{\lambda_0-\nu_0}\psi_1=0,
\end{equation}
where 
\begin{multline}
V[x]:=\mathrm{e}^{\nu_0-\lambda_0}\bigl\{
\tfrac{1}{2x}\partial_x[\nu_0-\lambda_0]
-\tfrac{1}{2}(\partial_x[\psi_0])^2
-\tfrac{x}{4}(\partial_x[\psi_0])^2\partial_x[\nu_0-\lambda_0]\\
+\tfrac{2}{x}\mathrm{e}^{\lambda_0}\sin(\psi_0)\partial_x[\psi_0]
+\tfrac{2}{x^2}\mathrm{e}^{\lambda_0}\cos(\psi_0)
+\tilde{\mu}^2\mathrm{e}^{\lambda_0}\cos(2\psi_0)
+\tfrac{1}{2}\tilde{\mu}^2x\mathrm{e}^{\lambda_0}\sin(2\psi_0)
\partial_x[\psi_0]\bigr\}.
\end{multline}
\end{subequations}

\begin{figure}[h]
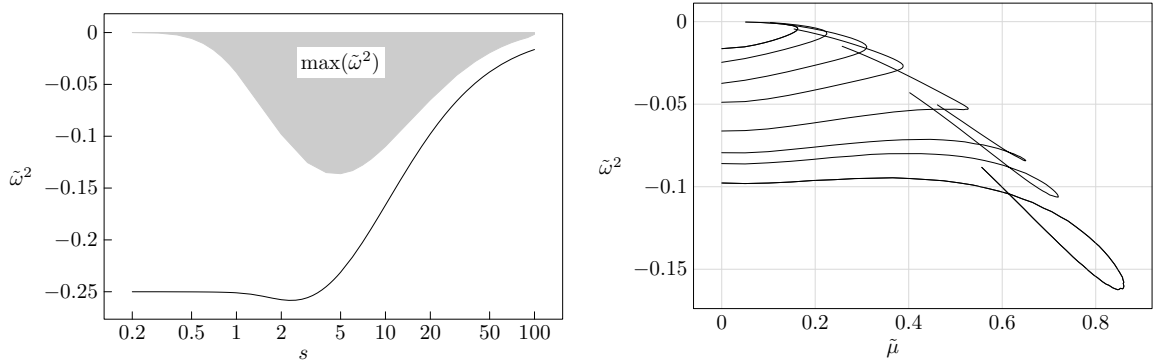

\begin{center}
\begin{subfigure}[$\tilde{\mu}=0$: The solid line represents the energy of the unstable mode determined numerically from the perturbation equation~\eqref{eq:pert eqns}, and the shaded region the upper bound determined from the variational method~\eqref{eq:var bound}]
{\includegraphics[scale=0.8]{Omega.ps}
\label{fig:omega 1}}
\end{subfigure}
\begin{subfigure}[$\tilde{\mu}\neq 0$:
The energy of the unstable mode for (bottom to top) $s=20$, $23$, $25$, $30$, $40$, $50$, $70$, $100$]
{\includegraphics[scale=0.8]{Omega-mu.ps}
\label{fig:omega 2}}
\end{subfigure}
\end{center}
\caption{
Unstable mode energy for $\mu=0$ and $\mu\neq 0$ solutions respectively.
}
\label{fig:omega}
\end{figure}

From the small-$x$ behaviour of the background fields~\eqref{eq:small} we find
\begin{equation}
\partial_x^2[\psi_1]-(1/x)\partial_x[\psi_1]+(1/x^2)\psi_1\sim 0,
\end{equation}
from which we have that $\psi_1\sim x(A+B\ln(x))$.
We will only consider cases where $B=0$ (corresponding to $\tau=0$ below) and the perturbation is therefore finite at $x=0$.
Once the (arbitrary) amplitude $A$ is chosen,~\eqref{eq:psi 1} is integrated numerically, varying the value of $\tilde{\omega}^2$ and requiring that $\psi_1\rightarrow 0$ as $x\rightarrow \infty$.
In Figure~\ref{fig:omega} we show the results of this procedure, giving the value of $\tilde{\omega}^2$ for the $\tilde{\mu}=0$ solutions, as well as a selection of $\tilde{\mu}\neq 0$ solutions.
In addition, we were unable to find any additional unstable modes.

\subsection{Variational Results}
\label{sect:variation}

Following Jetzer and Scialom~\cite{Jetzer+Scialom:1992} we introduce a radial coordinate $\rho$ that eliminates the second term in~\eqref{eq:psi 1}
\begin{subequations}
\begin{equation}
\rho=\int_0^x dx\;\mathrm{e}^{-\tfrac{1}{2}(\nu_0-\lambda_0)},
\end{equation}
chosen so that $\rho=0$ at $x=0$.
 From~\eqref{eq:small} and~\eqref{eq:asmpt Sch} respectively we find that
\begin{equation}
\rho\sim 
\begin{cases}
\sqrt{G/F} x^2/(a_p-1)&\text{as }x\rightarrow 0\\
x-\ln(x)+\mathrm{constant} &\text{as }x\rightarrow \infty\\
\end{cases};
\end{equation}
\end{subequations}
$\rho$ is determined numerically using the $x\rightarrow 0$ result and integrating outward.

This coordinate transforms~\eqref{eq:psi 1} to the stationary Schr\"{o}dinger-like equation
\begin{equation}\label{eq:Sch 1}
-\frac{\partial^2\psi_1}{\partial\rho^2}+V[x(\rho)]\psi_1=\tilde{\omega}^2\psi_1.
\end{equation}
We will consider perturbations in the Hilbert space $\mathcal{L}^2\bigl(d\rho,(0,\infty)\bigr)$,
\textit{i.e.}, with the norm
\begin{equation}
(\phi,\phi^\prime)=\int_0^\infty d\rho\,
\phi(\rho) \phi^\prime(\rho)
=\int_0^\infty dx\,
\mathrm{e}^{-\tfrac{1}{2}(\nu_0-\lambda_0)}
\phi[\rho(x)] \phi^\prime[\rho(x)].
\end{equation}
From the asymptotic and small $x$ forms given earlier, we find
\begin{equation}\label{eq:asympt V}
V[\rho]\xrightarrow{\rho\rightarrow 0}-1/(4\rho^2),\quad
V[\rho]\xrightarrow{\rho\rightarrow\infty}
\begin{cases}
2/\rho^2&\text{for }\tilde{\mu}= 0\\
\tilde{\mu}^2&\text{for }\tilde{\mu}\neq 0
\end{cases},
\end{equation}
and we therefore introduce
\begin{equation}
H_0=-\partial_\rho^2-\frac{1}{4\rho^2},\quad
\tilde{V}[x(\rho)]=\frac{1}{4\rho^2}+V[x(\rho)],
\end{equation}
so that~\eqref{eq:Sch 1} is written as
\begin{equation}\label{eq:Sch 2}
H_0\psi_1+\tilde{V}[x(\rho)]\psi_1=\tilde{\omega}^2\psi_1.
\end{equation}

From Narnhofer~\cite{Narnhofer:1974} and following~\cite{Jetzer+Scialom:1992}, we know that the operator $H_0$ is not self-adjoint on the intersecting domain $\mathcal{D}\bigl(-\partial_\rho^2\bigr)\cap\mathcal{D}\bigl(-1/(4\rho^2)\bigr)$, with deficiency indices $(1,1)$.
A family of self-adjoint extensions is determined by extending this domain to include the solutions of 
\begin{subequations}
\begin{equation}
-\partial_\rho^2[\psi_{\pm i}]-1/(4\rho^2)\psi_{\pm i}=\pm i\psi_{\pm i}, 
\end{equation}
which in this case are 
\begin{equation}
\psi_i=\sqrt{\rho}H^{(1)}_0\bigl(\rho\mathrm{e}^{i\pi/4}\bigr),\quad
\psi_{-i}=\bar{\psi}_i=\sqrt{\rho}H^{(2)}_0\bigl(\rho\mathrm{e}^{-i\pi/4}\bigr),
\end{equation}
\end{subequations}
where $H_0^{(1)}$ and $H_0^{(2)}$ are the zeroth-order Hankel functions of the first and second kind.

The real self-adjoint extensions are parameterized by the real angle $\tau$, and extend the operator to
act on the real functions
\begin{equation}\label{eq:Psitau}
\Psi_\tau:=\tfrac{1}{2}
\bigl(\mathrm{e}^{i\tau}\psi_i+\mathrm{e}^{-i\tau}\psi_{-i}\bigr)
=\cos(\tau)\Re(\psi_i)-\sin(\tau)\Im(\psi_i),
\end{equation}
where $\Re$ and $\Im$ represent the real and imaginary parts respectively.
We write the extended operator $\bar{H}_{0,\tau}$ which acts like $H_0$ on the extended domain
\begin{equation}
\mathcal{D}(\bar{H}_{0,\tau})
=\mathcal{D}\bigl(-\partial_\rho^2\bigr)\cap
\mathcal{D}\bigl(-1/(4\rho^2)\bigr)
+\{\Psi_\tau \},
\end{equation}
and acts on $\Psi_\tau$ as
\begin{equation}\label{eq:HPsitau}
\bar{H}_{0,\tau}\Psi_\tau=\tfrac{1}{2}i
\bigl(\mathrm{e}^{i\tau}\psi_i-\mathrm{e}^{-i\tau}\psi_{-i}\bigr)
=-\sin(\tau)\Re(\psi_i)-\cos(\tau)\Im(\psi_i).
\end{equation}
We will require the following integrals:
\begin{subequations}\label{eq:params}
\begin{gather}
\int_0^\infty d\rho\;\bigl(\Re(\psi_i)\bigr)^2
=\int_0^\infty d\rho\;\bigl(\Im(\psi_i)\bigr)^2
=:a\approx 0.1592,\\
\int_0^\infty d\rho\;\Re(\psi_i)\Im(\psi_i)
=:b\approx -0.1013,
\end{gather}
\end{subequations}
and using~\eqref{eq:Psitau} and~\eqref{eq:HPsitau} we find
\begin{equation}\label{eq:inn results}
(\Psi_\tau,\Psi_\tau)
=\bigl(a-b\sin(2\tau)\bigr)>0 ,\quad
(\Psi_\tau,\bar{H}_{0,\tau}\Psi_\tau)
=-b\cos(2\tau)\in[-b,b].
\end{equation}

From the large-$z$ behaviour of the Hankel function $H^{(1)}_0(z)\sim
\sqrt{\frac{2}{z\pi}}\exp\bigl(i(z-\pi/4)\bigr)+o\bigl(z^{-3/2}\bigr)$, we find the asymptotic behaviour of $\Psi_\tau$ for large $\rho$
\begin{equation}\label{eq:largez}
\Psi_\tau\sim
\sqrt{\frac{2}{\pi}}\mathrm{e}^{-\rho/\sqrt{2}}
\sin\bigl(\rho/\sqrt{2}+\pi/8+\tau\bigr),
\end{equation}
and from the small-$z$ behaviour of the Hankel function ($\gamma_{\textsc{e}}$ is Euler's constant) $H^{(1)}_0(z)\sim
1+\frac{2i}{\pi}\bigl(\ln(z/2)+\gamma_{\textsc{e}}\bigr)+o\bigl(z^2\ln(z)\bigr)$, we find near $\rho=0$ that 
\begin{equation}
\Psi_\tau\sim \sqrt{\rho}\Bigl(
\frac{1}{2}\cos(\tau)-\frac{2\gamma_{\textsc{e}}}{\pi}\sin(\tau)
-\frac{2}{\pi}\sin(\tau)\ln(\rho/2)\Bigr).
\end{equation}
At this point we note that only the $\tau=0$ extension (chosen implicitly in~\cite{Jetzer+Scialom:1992}) corresponds to a perturbative field that is finite at $\rho=0$ and we will restrict ourselves to this case in what remains.

\begin{figure}[h]
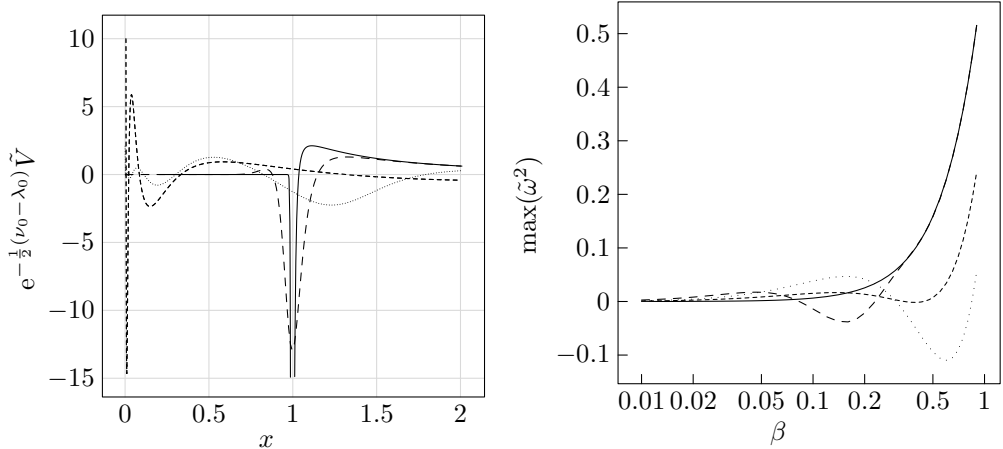

\begin{center}
\begin{subfigure}
{\includegraphics[scale=1.0]{tildeV.ps}}
\end{subfigure}
\begin{subfigure}
{\includegraphics[scale=1.0]{scanOmega.ps}}
\end{subfigure}
\end{center}
\caption{The effective potential and behaviour of $\mathrm{max}(\tilde{\omega}^2)$~\eqref{eq:var bound} with respect to $\beta$; the same values of $s$ as in Figure~\ref{fig:psi} are given.}
\label{fig:Vx}
\end{figure}

From the form of the differential operator $\bar{H}_{0,0}$, for any functions $\psi,\psi^\prime \in\mathcal{D}(\bar{H}_{0,0})$ we define $\psi_\beta(\rho):=\psi(\beta\rho)$, and it is straightforward to prove the scaling property: $(\psi^\prime_\beta,\bar{H}_{0,0}\psi_\beta)=\beta(\psi^\prime,\bar{H}_{0,0}\psi)$.
We therefore consider the collection of
\begin{equation}
\Psi_{0,\beta}(\rho):=\Psi_0(\beta\rho),
\end{equation}
as the variational family, varying $\beta$ to get an upper bound on the ground state energy.
Using this scaling property and the results~\eqref{eq:params} and~\eqref{eq:inn results}, we find
\begin{equation}
(\Psi_{0,\beta},\bar{H}_{0,0}\Psi_{0,\beta})
=-b\beta,\quad
(\Psi_{0,\beta},\Psi_{0,\beta})
=a/\beta,
\end{equation}
and taking the expectation value of~\eqref{eq:Sch 2} leads to the bound
\begin{equation}\label{eq:var bound}
\tilde{\omega}^2\leq
\beta\bigl((\Psi_{0,\beta},\tilde{V}\Psi_{0,\beta})-b\beta\bigr)/a
=:\mathrm{max}(\tilde{\omega}^2).
\end{equation}
Na\"{\i}vely we expect that very small values of $\beta$ will result in a bound of $\tilde{\omega}^2\le \tilde{\mu}^2$ since in these cases the bulk of the support of $\Psi_{0,\beta}$ is shifted to large $\rho$ and the asymptotic behaviour of the $\tilde{V}$~\eqref{eq:asympt V} in the expectation value dominates the integral.
In calculating the expectation value of the potential, we compute in practice
\begin{equation}
(\Psi_{\tau,\beta},\tilde{V}\Psi_{\tau,\beta})
=\int_0^{x_\mathrm{max}} dx\;\mathrm{e}^{-\tfrac{1}{2}(\nu_0-\lambda_0)}
\tilde{V}(x)\Psi^2_{\tau,\beta}[\rho(x)].
\end{equation}
The effective potential is plotted in Figure~\ref{fig:Vx} for some of the $\mu=0$ solutions, as is the dependence of $\mathrm{max}(\tilde{\omega}^2)$ on $\beta$.
We give the bound determined by minimizing $\mathrm{max}(\tilde{\omega}^2)$ with respect to $\beta$ in Figure~\ref{fig:omega 1}.
Although we will not be quantitative on this bound derived on the $\tilde{\mu}\neq 0$ solutions we have checked that it is consistent with the numerically determined $\tilde{\omega}^2$, generically providing a negative or zero bound.

\section{Discussion}

In this work we have examined the Wyman sector of the nonsymmetric gravitational theory in some detail.
We began by mapping it onto an Einstein-Klein Gordon system with metric $\hat{\mathrm{g}}$ and scalar field $\psi$ with a positive semi-definite (coordinate-dependent) interaction potential.
From this form we were able to employ a variational argument to show that there were neither globally regular solutions with nontrivial $\psi$ nor spacetimes with a horizon and nonzero $\psi$ in the exterior.
Considering small $r$ behaviour that matches the analytically known Wyman solution, we numerically solved the static background equations reproducing the analytically known Wyman solution and generating new solutions to the $\mu\neq 0$ field equations.

The dynamical stability of these solutions was investigated by direct numerical determination of the lowest energy state and corresponding eigenvalue (similar to that appearing in~\cite{Choptuik+Hirschmann+Liebling:1997}), as well as the variational approach of Jetzer and Scialom~\cite{Jetzer+Scialom:1992}.
We found that the $\mu=0$ Wyman solutions are generically unstable and therefore not possible candidates for the endpoint of a collapsing NGT system, however they are most likely threshold solutions (no evidence of more than one unstable mode was found) and may therefore play a role in critical collapse~\cite{Hara+Koike+Adachi:1997}.

\section*{Acknowledgements}

The authors thank the Natural Sciences and Engineering Research Council of Canada, the Walter~D.~Sumner Foundation, the Government of Ontario (OGS), and the Department of Physics at the University of Toronto for financial support.

\appendix

\section{The Wyman Solution}
\label{sect:Wyman}

Here we develop some analytic results from the Wyman solution~\cite{Bonnor:1951} of the static, $\mu=0$ field equations.
Adopting the form of the solution given in~\cite{Legare+Moffat:1995}, we have (the coordinates are labelled by $\{t,\sigma,\theta,\phi\}$; the functions $\gamma$, $\alpha$, $\beta$ and $\kappa$ are functions of the dimensionless radial coordinate $\sigma$ only)
\begin{subequations}
\begin{equation}\label{eq:Wyman}
\lvert\mathrm{g}_{(\mu\nu)}\rvert=\mathrm{diag}
\bigl(\gamma,-\alpha,-\beta,-\beta\sin^2(\theta)\bigr),\quad
\mathrm{g}_{[23]}=\kappa\sin(\theta),
%
%
\end{equation}
where $\sigma\in[0,\infty]$, and we consider two separate solutions (corresponding to the $\pm$ signs throughout)
\begin{align}
\gamma(\sigma)=&\mathrm{e}^{\pm\sigma},\\
\alpha(\sigma)=&4M_{\textsc{s}}^2\mathrm{e}^{\mp\sigma}(1+s^2)C(\sigma)^{-2},\\
\beta(\sigma)=&8M_{\textsc{s}}^2\mathrm{e}^{\mp\sigma}C(\sigma)^{-2}
\bigl(\cosh(a_p\sigma)\cos(a_m\sigma)-1+s\sinh(a_p\sigma)\sin(a_m\sigma)\bigr),\\
\kappa(\sigma)=&8M_{\textsc{s}}^2\mathrm{e}^{\mp\sigma}C(\sigma)^{-2}
\bigl(\sinh(a_p\sigma)\sin(a_m\sigma)+s-s\cosh(a_p\sigma)\cos(a_m\sigma)\bigr).
\end{align}
\end{subequations}
Here 
\begin{equation}
a_p:=\sqrt{\frac{\sqrt{1+s^2}+1}{2}},\quad 
a_m:=\sqrt{\frac{\sqrt{1+s^2}-1}{2}},
\end{equation}
$s$ is a dimensionless constant of integration, and $C(\sigma)$ is defined by
\begin{equation}
C(\sigma):=2\bigl(\cosh(a_p\sigma)-\cos(a_m\sigma)\bigr).
\end{equation}
From its definition $C(\sigma)$ and its derivative are non-negative, strictly increasing functions of $\sigma$, and furthermore $\partial_\sigma[C(\sigma)]\ge C(\sigma)$.

We use the dimensionless coordinate $x:=r/(2M_{\textsc{s}})$ where $r$ is defined by requiring that the fundamental tensor be of the form~\eqref{eq:ft param}, so that $\beta^2+\kappa^2=r^4$ and we find
\begin{equation}\label{eq:r rel}
x_s^2=\mathrm{e}^{\mp\sigma}/C(\sigma),\quad
\text{and}\quad
\partial_\sigma[x]=-\tfrac{1}{2}x\bigl
(\partial_\sigma[\ln C(\sigma)]\pm 1\bigr)\le 0.
\end{equation}
For notational simplicity we have also introduced
\begin{equation}
x_s:=x/(1+s^2)^{1/4}.
\end{equation}
Performing this coordinate transformation, we find the components of~\eqref{eq:ft param}
\begin{equation}
\mathrm{e}^{\nu}=\mathrm{e}^{\pm\sigma},\quad
\mathrm{e}^{\lambda}:=\alpha(\sigma)(\partial_r[\sigma])^2
=\frac{4\sqrt{1+s^2}C(\sigma)}
{\bigl(\partial_\sigma[C(\sigma)]\pm C(\sigma)\bigr)^2},
\end{equation}
and $\psi$ may be determined from
\begin{equation}\label{eq:tanpsi}
\tan(\psi)=\frac{\sinh(a_p\sigma)\sin(a_m\sigma)
+s\bigl(1-\cosh(a_p\sigma)\cos(a_m\sigma)\bigr)}
{s\sinh(a_p\sigma)\sin(a_m\sigma)
-\bigl(1-\cosh(a_p\sigma)\cos(a_m\sigma)\bigr)}.
\end{equation}

For small $\sigma$ the relationship~\eqref{eq:r rel} is easily inverted to give $\sigma\sim 1/x$,
which leads to $C(\sigma)\sim\sqrt{1+s^2}\sigma^2\sim 1/x_s^2$ and 
\begin{equation}\label{eq:large x}
\mathrm{e}^{\nu}\sim 1\pm 1/x,\quad 
\mathrm{e}^{\lambda}\sim 1\mp 1/x,\quad
\psi\sim s/(12x^2),
\end{equation}
justifying the identification of $M_{\textsc{s}}$ as the asymptotic Schwarzschild mass
parameter with the lower (upper) sign corresponding to the positive (negative) mass solution (in
deriving the final result in~\eqref{eq:large x} the numerator in~\eqref{eq:tanpsi} must be
expanded to fourth-order in $\sigma$).
These results are also consistent with the $s\rightarrow 0$ limit of~\eqref{eq:r rel}, from which one finds $\mathrm{e}^{\nu}=(1\pm 1/x)$, and we find the negative and positive mass Schwarzschild solutions respectively (only the exterior in the latter case since the radial coordinate only extends to $r=2M_{\textsc{s}}$ in the limit).

For large $\sigma$ we find $x_s\sim \exp\bigl(-(a_p\pm 1)\sigma/2\bigr)$, which may be inverted ($s\neq 0$) to give 
\begin{equation}
\sigma\sim -\frac{2}{a_p\pm 1}\ln(x_s),\quad
C(\sigma)\sim\mathrm{e}^{a_p\sigma}
\sim x_s^{-2a_p/(a_p\pm 1)},
\end{equation}
and from these we find
\begin{equation}\label{eq:small x}
\mathrm{e}^{\nu}\sim 
x_s^{\mp 2/(a_p\pm 1)},\quad
\mathrm{e}^{\lambda}\sim\frac{4\sqrt{1+s^2}}{(a_p\pm1)^2}
x_s^{2a_p/(a_p\pm 1)}.
\end{equation}
Note that for the positive mass solution $\mathrm{e}^{\nu}\rightarrow 0$ as $x\rightarrow 0$ whereas for the negative mass solution $\mathrm{e}^{\nu}$ is singular as $x\rightarrow 0$; note that neither case satisfies the boundary conditions~\eqref{eq:bcs} and are therefore not covered by the scaling argument of Section~\ref{sect:background}.

Deriving the small $x$ behaviour of $\psi$ requires slightly more work.
First note that $\partial_\sigma[\psi]=-\partial_\sigma[\cos(\psi)]/\sin(\psi)$ and we have
\begin{subequations}
\begin{align}
\cos(\psi)=\frac{2}{\sqrt{1+s^2}}C^{-1}(\sigma)
\bigl(\cosh(a_p\sigma)\cos(a_m\sigma)-1+s\sinh(a_p\sigma)\sin(a_m\sigma)\bigr),\\
\sin(\psi)=\frac{2}{\sqrt{1+s^2}}C^{-1}(\sigma)
\bigl(\sinh(a_p\sigma)\sin(a_m\sigma)+s-s\cosh(a_p\sigma)\cos(a_m\sigma)\bigr),
\end{align}
\end{subequations}
from which we find in the large $\sigma$ limit that $\partial_\sigma[\psi]\sim a_m$.
Integrating this and re-writing it in terms of $x$ gives
\begin{equation}\label{eq:psi small x}
\psi\sim -\frac{2a_m}{a_p\pm 1}
\ln(x_s)+\tan^{-1}(-s).
\end{equation}
The constant is determined by noting that for $\sigma=\sigma_n:=n\pi/a_m$, $n=1,2,\ldots$
($n=0$ is a special case since both the numerator and denominator of~\eqref{eq:tanpsi} become
degenerate) we have from~\eqref{eq:tanpsi} that $\tan(\psi)=-s$, and therefore
\begin{equation}\label{eq:points}
\psi(\sigma_n)=n\pi +\tan^{-1}(-s).
\end{equation}
As a check on the accuracy of the numerical solutions, the first six of these points are indicated by diamonds on the numerically generated solutions in Figure~\ref{fig:psi}.

Note that for this solution, the components of the metric $\hat{\mathrm{g}}$ as defined in~\eqref{eq:ghat} are well-behaved except at $r=0$ where $\hat{\mathrm{g}}$ has a curvature singularity which shows up in the Ricci scalar~\eqref{eq:hat equations} as 
\begin{equation}
\hat{R}\sim -\frac{2}{r^2}
\Bigl(1+\frac{a_m^2}{4\sqrt{1+s^2}} x_s^{-2a_p/(a_p-1)}\Bigr).
\end{equation}
Furthermore, by considering radial null geodesics ending at $r=0$ we find that this is a strong curvature singularity.
Radial null geodesics affinely parameterised by $\lambda$ are determined from the tangent vector ($E$ is a constant of integration) $u=\bigl(\partial_\lambda[t],\partial_\lambda[r]\bigr)=\bigl(E\mathrm{e}^{-\nu},\pm E\mathrm{e}^{-(\nu+\lambda)/2}\bigr)$ which may be integrated near $r=0$ to find (choosing $\lambda=0$ at $r=0$, $\lambda$ increases with $r$ and assuming that the trajectory passes through $r=0$ at $t=0$) $r^2=(a_p\pm1)2M_{\textsc{s}}t$ and $r\sim 2M_{\textsc{s}}(1+s^2)^{1/4}\bigl(a_pE\lambda/(2M_{\textsc{s}}\sqrt{1+s^2})\bigr)^{(a_p\pm 1)/(2a_p)}$.
Using the result that $\hat{R}(u,u)=-E^2\partial_r\bigl[\mathrm{e}^{-(\nu+\lambda)}\bigr]/r$, we compute the derivative and write what remains in terms of the affine parameter to find $\lambda^2\hat{R}(u,u)\sim a_m^2/(2a_p^2)$; the finite limit of this as $\lambda\rightarrow 0$ indicates that $\hat{\mathrm{g}}$ has a strong curvature singularity at $r=0$ by an argument of Clarke and Krolak (see for instance~\cite{Tipler+Clarke+Ellis:1980}).

Outgoing null geodesics are determined from $\partial_\lambda[r]=E\mathrm{e}^{-(\nu+\lambda)/2}$, and upon integrating from $r=0$ (fixing $\lambda=0$ at $r=0$) and rewriting it in terms of $\sigma$ we find that 
\begin{equation}
\int_0^r dr\,\sqrt{\alpha\gamma}
=2M_{\textsc{s}}\sqrt{1+s^2}\int_{\sigma_0}^\infty d\sigma\,C(\sigma)^{-1}=E\lambda.
\end{equation}
This integral is strongly convergent as $\sigma_0\rightarrow \infty$, and has no poles except at $\sigma=0$ where the integral diverges.
This indicates that the affine parameter $\lambda$ only goes to infinity as $\sigma_0\rightarrow 0$ ($r\rightarrow \infty)$; hence outgoing radial null geodesics reach any finite radius in a finite affine time, and the singularity is therefore visible (naked).
Identical results may be derived from the so-called path equation ({\it c.f.},~\cite{Legare+Moffat:1995b}, Section~5, in particular~(24)).



\end{document}